\documentclass[graybox]{svmult}

\usepackage[colorlinks,citecolor=blue,linkcolor=blue,urlcolor=blue]{hyperref}

\usepackage{color}
\usepackage{amsmath,amssymb}

\usepackage{mathptmx}       
\usepackage{helvet}         
\usepackage{courier}        
\usepackage{type1cm}        
\usepackage{makeidx}         
\usepackage{graphicx}        
\usepackage{multicol}        
\usepackage[bottom]{footmisc}
\usepackage{graphicx,subfigure}
\begin{document}

\title*{The Good, the Bad, and the Ugly of Gravity and Information }
 \titlerunning{The Good, the Bad, and the Ugly}

\author{Gerard~'t Hooft, Steven~B. Giddings, Carlo Rovelli, Piero Nicolini, Jonas Mureika, Matthias Kaminski, and Marcus Bleicher}

\authorrunning{'t Hooft, Giddings, Rovelli, Nicolini, Mureika, Kaminski, Bleicher}

\institute{
Gerard~'t Hooft \at Institute for Theoretical Physics,
EMME$\Phi$, Centre for Extreme Matter and Emergent Phenomena, Science Faculty, Utrecht University,  
POBox 80.089, 3508 TB Utrecht,
The Netherlands, \email{g.thooft@uu.nl}
\and Steven~B. Giddings  \at Department of Physics, University of California, Santa Barbara, CA 93106, USA,
\email{giddings@physics.ucsb.edu}
\and Carlo Rovelli \at Aix Marseille Universit{\'e}, CNRS, CPT, UMR 7332, 13288 Marseille, France, and Universit{\'e} de Toulon, CNRS, CPT, UMR 7332, 83957 La Garde, France, 
\email{rovelli@cpt.univ-mrs.fr}
\and Piero Nicolini \at Frankfurt Institute for Advandced Studies (FIAS),
Ruth-Moufang-Str. 1, 60438 Frankfurt am Main, Germany, and Institut f{\"u}r Theoretische Physik, Johann Wolfgang Goethe-Universit{\"a}t Frankfurt am Main
Max-von-Laue-Str. 1, 60438 Frankfurt am Main, Germany,
\email{nicolini@fias.uni-frankfurt.de}
\and Jonas Mureika \at Department of Physics, Loyola Marymount University,
1 LMU Drive, Los Angeles, CA 90045, USA,
\email{jmureika@lmu.edu}
\and Matthias Kaminski \at Department of Physics and Astronomy, University of Alabama, Tuscaloosa, AL 35487, USA,
\email{mski@ua.edu}
\and Marcus Bleicher \at Frankfurt Institute for Advandced Studies (FIAS),
Ruth-Moufang-Str. 1, 60438 Frankfurt am Main, Germany, and Institut f{\"u}r Theoretische Physik, Johann Wolfgang Goethe-Universit{\"a}t Frankfurt am Main
Max-von-Laue-Str. 1, 60438 Frankfurt am Main, Germany,
\email{bleicher@fias.uni-frankfurt.de}
}
%
%
\maketitle
\vskip -3cm

\abstract{
Various contenders for a complete theory of quantum gravity are at odds with each other. This is in particular seen in the ways they relate to information and black holes, and how to effectively treat quantization of the background spacetime. 
Modern perspectives on black hole evaporation suggest that quantum gravity effects in the near-horizon region can perturb the local geometry.
The approaches differ, however, in the time scale on which one can expect these effects to become important. This panel session presents three points of view on these problems, and considers the ultimate prospect of observational tests in the near future.}

\begin{flushright}

\vskip -0.5cm
\end{flushright}
{\em 
    ``Two hundred thousand dollars is a lot of money. We're gonna have to earn it.''\\
\\
\vskip -0.5cm
\rightline{{\rm --- Blondie, ``The Good, the Bad and the Ugly'', 1966}}
}

\bigskip

\section{Introduction}

In 1975 Hawking obtained a ground breaking result about the fundamental nature of black holes \cite{Hawking:1974sw} that highlighted three crucial characteristics.  Firstly, quantum mechanics allows for particle emission from black holes. Secondly, the spectrum of such an emission is thermal\footnote{
The spectrum at the event horizon is thermal in an extremely accurate approximation according to Hawking's argument. Small corrections come from the decreasing black hole mass due to evaporation, as well as finite size and shape effects during the emission. An asymptotic observer measures devations from a thermal spectrum induced by the curved geometry outside the horizon, i.e. greybody factors.} 
in the sense of black body radiation. Lastly, the temperature of the emitted particles is proportional to the black hole's surface gravity. Although relations between horizon area, surface gravity, and black hole mass resembling the laws of thermodynamics were already known at that time \cite{Bardeen:1973gs}, the idea of black hole thermodynamics was only taken seriously after Hawking's derivation of  ``black hole evaporation''. The thermal nature of black holes has since stimulated an immense number of investigations, and more importantly, intersected several research fields, such as particle physics, cosmology, statistical physics, and information theory. Since for (asymptotically flat Schwarzschild) black holes\footnote{This is the system which we consider in this paper if not stated otherwise.} the temperatures increase as their masses decrease, soon after Hawking's discovery, it became clear that a complete description of the evaporation process would ultimately require a consistent quantum theory of gravity.  This is necessary as the semiclassical formulation of the emission process breaks down during the final stages of the evaporation as characterized by Planckian values of the temperature and spacetime curvature. 

More than 40 years after Hawking's discovery the situation remains unclear and a variety of issues unresolved.
A quantum theory of gravity is not only expected to provide an ultraviolet completion of general relativity, but also to describe consistently the microscopic degrees of freedom at the basis of the statistical interpretation of black hole thermodynamic variables   \cite{Chirco:2014saa,SGstatph,Jacobson:1995ab,Padmanabhan:2003gd,Padmanabhan:2009vy}. 
For instance, both the heat capacity and the entropy of black holes exhibit anomalous behavior. Black holes  have negative heat capacity throughout the entire evaporation process, while their  entropy is proportional to the area of the event horizon rather than the interior volume, as is the case in  classical thermodynamic systems. 
As acceptance of the area-entropy law as a realization of the holographic principle  
\cite{Bousso:2002ju,Maldacena:1997re,Susskind:1994vu,'tHooft:1993gx} is still debated between communities, also black hole thermodynamics still posits important issues related to our understanding of fundamental physics.
At the classical level, the formation of an event horizon as a result of a gravitational collapse implies information loss of the star's initial microstates. Quantum mechanical effects worsen the situation.  Thermal radiation is a mixed quantum mechanical state, and therefore black hole evaporation challenges one of the basic principles of quantum mechanics, \textit{i.e.}, the impossibility for pure states to evolve into such mixed states.

Several formulations have been proposed in order to address issues raised by black hole evaporation, which we term  the Good, the Bad, and the Ugly:

\begin{itemize}
\item {\bf The Good}: There exist some model-independent characteristics, or at least an agreement on how the semiclassical description of black hole thermodynamics has to be improved. For instance, the nature of the black hole entropy is often interpreted in terms of  entanglement entropy 
\cite{Bianchi:2014bma,SGstatph,McGough:2013gka}. 
But ultimately some modifications are expected in the vicinity of the horizon. It is expected that the usual notions of locality and causality will  be violated when both gravitational and quantum mechanical effects are simultaneously taken into account. These violations might allow information to  leak out of the horizon (see \textit{e.g.} \cite{Averin:2016ybl,BHMR,LQGST,SGmodels,Hawking:2016msc,thooft,Hooft:2016itl}).  
Traditionally, the above issues were approached by maintaining the dogmas of gravity at the expense of local quantum field theory principles. However, it is also possible to postulate the breakdown of gravity while maintaining quantum field theory. In this way, the possibility of a black hole firewall \cite{Almheiri:2012rt} might be considered. 

\item {\bf The Bad:} To date, we do not have any experimental/observational data in order to discriminate among the plethora of possible scenarios, even though there exists a new generation of facilities, \textit{e.g.} the Event Horizon Telescope \cite{EHT}, Advanced LIGO, \cite{ligo,Abbott:2016nmj}  as well as  future runs at the Large Hadron
Collider (LHC, scheduled until 2035) \cite{LHCRUNS}, that have the potential to disclose crucial clues. 

\item{\bf The Ugly:} We need to dig deep into a complicated mess of mathematical formalism to understand the relationship between gravity and information and to extrapolate reliable phenomenological scenarios.
\end{itemize}
In the following pages, we expand on these topics to show the ways in which they conflict, but also the ways in which they complement  one another.

\section{{The Loop Quantum Gravity Perspective}\\
\small{\it Contribution by C.~Rovelli}}

The convergence of ideas we have witnessed in this conference is surprising, and is good news. I was surprised how Steve and I, coming from different theoretical paths, have come to similar conclusions.  

The first convergence point is the growing conviction that quantum gravitational phenomena can violate the basic assumptions of standard Local Quantum Field Theory (LQFT).  They can violate the causality dictated by the background geometry.   On the other hand, I understand that Gerard disagrees with this, and does not expect quantum gravity to violate the LQFT basics.

There is a difference, however, between Steve's and my 
views on the violation of LQFT causality.  For Steve, this is a shocking new phenomenon that points to some mysterious new physics and demands some deep revision of current physics and some courageous new speculation.  For me --- in fact for a large community of people that have been working on quantum gravity for decades --- this is the natural consequence to be expected when general relativity and quantum theory are taken together.  

Let me explain: we have learned from general relativity that spacetime geometry --- therefore the causal structure --- are determined by the gravitational field.  The gravitational field is a quantum field, therefore the geometry it determines is not going to be sharp: it undergoes quantum fluctuations, can be in superpositions, and entangled.  Therefore the causal structure of a single fixed background geometry can be violated.

To have such violations, one must exit the regime of validity of perturbation theory, because perturbation theory can be formulated over a background configuration respecting the causality of the background.   But of course there are  phenomena in quantum mechanics that escape perturbation theory: quantum tunnelling is a prominent example. Therefore tunnelling phenomena can easily violate the causality of conventional LQFT, because LQFT does not allow a violation of the causality determined by the background geometry. 

What is the time scale for these violations?   There is a simple dimensional argument that gives an indication. To have quantum gravitational phenomena, you need something to get to Planckian scale, which is to say to get to order 1 in Planck units. One well known possibility is to have a very high curvature $R\sim 1$: this happens near the region where classical General Relativity (GR)
predicts a singularity, indicating that classical GR fails there.  But there is another possibility: small quantum corrections can pile up and give radical departures from classicality over a long time $T$.  Therefore  in a region of low curvature we might already see quantum phenomena after a time $T$, as soon as $RT\sim 1$.  Around a mass $m$ the curvature goes like $R\sim m/r^3$, where $r$ is the radius, and in the region outside the horizon of a black hole $r\sim m$. This gives 
$$
                     T\sim m^2 \, ,
$$
as a possible time to see quantum gravitational effects.  For a macroscopic hole, this is a huge time: it is the Hubble time for a millimeter-size black hole. 
But it is still enormously smaller than the immense Hawking evaporation time, which is $T\sim m^3$, meaning that it takes $10^{35}$ Hubble times to evaporate the same millimetre size black hole.   This indicates that for a black hole there can be \emph{other} quantum phenomena than the Hawking radiation, taking a shorter time than the Hawking time, which are not seen by LQFT. 

There is another way of seeing that quantum gravity allows for surprising violations of causality.   Suppose for a moment that causality could be violated at the distance of a single Planck length.  For instance, suppose information could be fast transmitted one Planck length away in a spacelike direction.  Take an arbitrary point $P$ in the classical metric of a collapsed star, sufficiently after the collapse.  It is a fact that the distance between $P$ and the singularity is smaller than a Planck length. This is completely counter-intuitive for a Newtonian intuition, but it is a simple consequence of special relativity: a light ray emitted from a point just before $P$ can reach the singularity, therefore there are points of the singularity at an arbitrarily 
small spacial distance from $P$.  If quantum effects can spread information a Planck distance away, they can move information from the singularity to anywhere in space.  No wild speculation about new physics is required for this: general relativity and quantum theory suffice. But we must go beyond LQFT. 

A simple possibility of  a quantum gravitational phenomenon that can happen as soon as classical causality does not constrain quantum gravity phenomena, is the following:  a macroscopic black hole can 
 ``explode", namely tunnell-out to a white-hole, while still macroscopic, without having to wait  for the end of Hawking's evaporation
\cite{
Ashtekar:2005cj,
Balasubramanian:2006,
Bambi2013,
Barcelo:2015uff, 
Bardeen2014,
Carr:2015nqa,
frolov:BHclosed,
Frolov:1979tu,
Frolov:1981mz,
Gambini:2013qf,
Giddings1992b,
HAJICEK2001,
Hayward2006,
Hossenfelder:2010a,
Isi:2013cxa,
Mathur2005,
Mathur2015,
Mazur:2004,
modesto2004disappearance,
Modesto2008a,
Narlikar1974,
Nicolini:2005vd,
Nicolini:2008aj,
Rovelli2014,
Saueressig2015,
Spallucci:2009zz,
Stephens1994}. 
This is akin to standard nuclear decay, which is a prototypical tunnelling phenomenon.  Remarkably, it has been shown in \cite{Haggard2014} that this is possible \emph{without} violating the classical equations of motion outside a compact spacetime region.  A black hole can thus quantum gravitationally tunnel into a white hole and explode.  This is a standard quantum tunneling 
phenomenon, and therefore there is no plausible reason for it not to happen.  The relevant physical question is how long it takes. If it takes too long, it is not of astrophysical relevance (but it \emph{still} shows that LQFT is going to be violated).  If, on the other hand, the dimensional estimate above ($T\sim m^2$) is correct, then this phenomenon can have astrophysical relevance because millimeter size primordial black holes could be exploding today, leading to observable cosmic rays. A millimeter size black hole has the mass of a planet. In the sudden explosion triggered by quantum gravitational tunneling the huge corresponding energy is projected out. 
It has been conjectured that some Fast Radio Bursts and high energy gamma rays could have this origin \cite{Barrau2014c,Barrau2014b}.

This is the second remarkable point of convergence between Steve and I:  we are now talking of potentially observable quantum gravity. We even have calculations trying to compute the distance from the horizon where violations of the classical could be expected \cite{SGGW,Haggard:2016ibp}. 
For a field long in search of observations \cite{Amelino-Camelia2013,Liberati2011}, this is again good news.   Maybe black holes could `reveal their inner secrets' \cite{Jacobson2010a} after all, thanks to quantum theory. 

Of course in order to describe this kind of phenomenon we need a formulation of quantum general relativity which is not in the form of a LQFT over a fixed geometry.  It is in this sense that we need something radical to understand quantum black holes.  That is, not in the sense that we need mysterious new physics; but  in the sense that we have to accept the idea that combining general relativity and quantum theory requires us to abandon the framework of LQFT. 

Loop quantum gravity \cite{Ashtekar:2012eb,Gambini,Rovelli:2004fk,Rovelli:2014ssa,ThiemannBook} does provide a formulation of quantum gravity which is background free, and it does indicate that non-perturbative phenomena violating the causality of the background geometry are possible.  Explicit calculations are in course to use LQG to compute the lifetime of a black hole under this kind of decay \cite{Christodoulou2016}.    In the bounce region, the `architecture' \cite{Bianchi2012b} of the quantum geometry is fully non-classical. 

I close with a physical picture of the causality violation. The simplest way to interpret black hole entropy is in terms of quantum field entanglement across the horizon. The traditional difficulty of this interpretation is the ``species problem", namely the naive expectation that the amount of entanglement entropy  should depend on the number of existing fields and not have the universal character of the Bekenstein-Hawking entropy.  This difficulty has been brilliantly solved by Bianchi, in \cite{Bianchi2012} by showing that $dS=dA/4$ is independent from the ultraviolet cut-off and from the number of species; it is an infrared phenomenon that follows simply from the Einstein equations and standard QFT.\footnote{As a side remark: the famous $1/4$ factor of the Bekenstein-Hawking entropy $S=A/4$ is confusing: if instead of $G_{\rm Newton}=1$ we use units where the proper coupling constant of GR is taken to be unit, namely $8\pi G_{\rm Newton}=1$, then the coefficient of the Bekenstein-Hawking entropy looks far more conventional: $S=2\pi\, A$.} Now, the interpretation of black hole entropy as entanglement entropy may seem to support the solidity of a physical picture where the background geometry can be considered fixed.  But this would be wrong: among the entangled fields is the gravitational field itself, which means that the geometry on the horizon fluctuates, which means that the causal structure on the horizon fluctuates.  This is a physical process allowing information to escape, of course. Preliminary calculations in Loop Quantum Gravity \cite{Chistodoulou2016} show that the amplitudes for black hole explosion come precisely from interference between different eigenstates of the horizon geometry.

Overall, I feel that we are making excellent progress in understanding quantum gravity and quantum black holes.  What restrains  understanding  is excessive trust in the validity of LQFT for describing quantum gravitational physics.  This is still widely diffused among theoretical physicists. 

\section{Black Holes and Correspondence as Guides to New Principles\\
\small{\it Contribution by S.~Giddings}}

In my talk I outlined arguments for important quantum modifications to a description of black holes that has historically been based on perturbative local quantum field theory (LQFT) on a semiclassical background, which for example might be closely approximated by Schwarzschild's original solution.  Here, I'll review some of these arguments and respond to some of the comments made by others in our discussion, and also comment further on the prospects for observation of or observational constraints on black hole quantum structure.  This  will summarize more in-depth discussion from a series of papers I've written; references will be given to guide the reader to more detailed arguments.
\subsection{Quantum modifications to standard locality}

There has been a growing sense in the theoretical community that the only way we can consistently describe black hole evolution is to accept that there are significant modifications to the treatment based on perturbative quantization of matter and metric fluctuations on a background geometry, which can be semiclassically corrected to account for Hawking flux.  In particular, a central issue appears to be the role of locality in quantum gravity; it is the usual locality of LQFT that forbids transfer of information (signaling) from the interior of a black hole to its exterior, but transfer of information from black hole ``internal" states to exterior is apparently exactly what is ultimately needed for a unitary description of evolution.  An important point is that while everyone expects a breakdown of locality at very short ({\it e.g.} Planckian) scales, we have now realized that what is apparently needed is a modification of the LQFT description of locality at {\it long} distance scales -- comparable to the horizon radius scale of even very large black holes.  It is hard to see how short-distance modifications to locality yield this result.

Of course the question of locality in quantum gravity is a tricky one.  Indeed, in gravity there is an obstacle to formulating local gauge (diffeomorphism)-invariant observables; such local observables are used in non-gravitational LQFT to sharply characterize locality (see, {\it e.g.}, \cite{Haag}).  As has been made even more  precise recently \cite{DoGi,Donnelly:2016rvo,SGalg}, 
the ``gravitational  dressing" required to satisfy the gravitational constraint equations, or their quantum version, produces an obstacle to this LQFT locality.  

There has been a lot of discussion over time whether effects related to this observation actually save us in the black hole context.  For example, the proposal that black holes carry quantum gravitational hair~\cite{Hawking:2016msc} appears closely related, as do 
Gerard's 
suggestions in the discussion that gravitational backreaction communicates the needed quantum information from ingoing particles to outgoing Hawking particles~\cite{thooft}. However, so far it has been very hard for many of us to see that such effects can be strong enough to restore unitarity to black hole evaporation; 
certain toy models for black hole evaporation, {\it e.g.} \cite{CGHS}, seem to reinforce the case that gravitational dressing or backreaction is not enough.\footnote{{Note  for example that there is a cancellation between the interaction of infalling matter with outgoing Hawking particles and with Hawking ``partners" behind the horizon~\cite{Giddings:2006sj}, which appears to eliminate large effects like those proposed by Gerard.}}    
I would make the same comment in response to Carlo's suggestion that 
fluctuations in the metric and causal structure are sufficient -- certainly there are perturbative studies of such fluctuations, and while there {have been} arguments that these can become important~\cite{SG2007} {at long times}, 
I don't know 
of clear arguments that their proper treatment can achieve the needed unitarization of  Hawking's original story by sufficiently delocalizing information or effectively transferring information from black hole states to the black hole exterior.

So in short, given the profound conflict we have encountered between the principles of relativity, of quantum mechanics, and of locality in describing black holes, and the ensuing ``unitarity crisis,"  
I am proposing 
that we {need to} consider fundamentally new quantum effects that do not respect the locality principle as formulated in LQFT on a semiclassical geometry, and that do not arise from a na\"ive quantization of general relativity.  Such effects seem necessary, in order to save unitarity and quantum mechanics in the black hole context.
I don't see an easier way out of our quandry, and this explains the origin of my talk title, ``Beyond Schwarzschild."

In fact, here we encounter the ``ugly" of the story: we don't presently seem to have a set of foundational principles to describe quantum gravity.  Of course many string theorists believe that AdS/CFT or related dualities provide such fundamental formulation, but many puzzles about how this could work remain, and skepticism~\cite{IsIt} has grown in the community.  My own point of view is that we need to think more generally, but that the ``good" includes the statement that  at least quantum mechanics, in a suitably general formulation (see, {\it e.g.}, \cite{UQM}), should be an essential element of the foundation of the theory.  Another part of the ``good" is the notion of correspondence for quantum gravity~\cite{SGnonpert,Erice,SGalg} -- whatever the more fundamental formulation is, it should match on to LQFT on semiclassical spacetime in the context of weak gravitational fields.  This should be an important guide helping us to infer the necessary additional mathematical structure that is needed beyond basic quantum postulates such as the existence of a Hilbert space.  
Quantum mechanics and correspondence together appear to be very powerful constraints.

In the absence of the complete fundamental structure, but assuming that it fits within the framework of quantum mechanics,
I have  taken 
the pragmatic viewpoint that we need to model and parameterize the relevant dynamics; doing so is in turn hoped to furnish important clues about this fundamental structure.  Specifically, since the perturbative gravitational dressing appears to be a ``weak" effect, 
I'll begin with 
the assumption that at the least we have a localization of information~\cite{BHQIUE} into quantum subsystems such as ``black hole," ``black hole atmosphere," and ``asymptotic spacetime;" far from a black hole we expect this localization of information to be described, to a very good approximation, just like in LQFT.  Then, we can investigate the  kinds of interactions between these subsystems that are needed to provide a unitary description of evolution.\footnote{Note that if nonlocality from gravitational dressing indeed is found to play a central role, its effect might also possibly be parameterizable in such a fashion.}
 
 In such a subsystem picture, the problem is that the 
Hawking process builds up entanglement between the black hole and the exterior, as is easily seen in a description where entangled Hawking particles and their partners are produced, with the latter residing inside the black hole.  It is this growth of entanglement that ultimately contradicts unitarity, once a black hole evaporates completely; to avoid this, the entanglement with the black hole states must be transferred out~\cite{SGmodels,GiShone,Sussxfer}  as the black hole decays.  Here is where the novel effects, which do not respect locality of LQFT, are needed.

I will also make what I consider to be 
the reasonable assumption that we should look for the ``minimal," or most conservative, departures from the usual story of LQFT.  So, while transfer of information from the black hole states to the exterior is needed, we might expect this information transfers to the states in the immediate vicinity of the black hole, and not, for example, to states  at a million times the Schwarzschild radius $R_S$.  Likewise, while we could consider drastic departures from the semiclassical spacetime geometry near the horizon, 
I will 
explore the assumption that we seek physics that produces minimal such departure, so, for example, an observer could still sail into a large enough black hole without being harmed at the point where he or she would expect to be crossing the horizon.

\begin{figure}
\hspace{-2cm}\includegraphics[height=10cm]{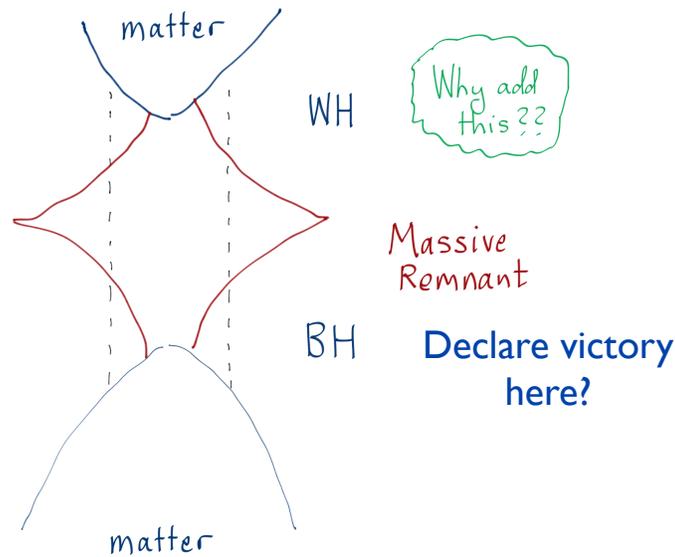}
\caption{\label{SGfig1}
{A spacetime diagram corresponding to the scenario Carlo described in his talk.  The Planckian region crosses the black hole horizon in a spacelike, ``nonlocal," fashion, like in the massive remnant scenario of \cite{BHMR}.  Once this is allowed, in principle we can describe information transfer from inside the black hole to outside.  Carlo also symmetrically adds a white hole piece to his spacetime.}}
\end{figure}

Here, by the way, I have evolved in my thinking.  
When I first proposed 
\cite{BHMR} that transfer of information that is nonlocal with respect to the semiclassical geometry is the way out of this crisis, {this was part of a} ``violent" picture in which the information escapes through formation of a new kind of object -- a massive remnant, which provides a sharp interface to the exterior vacuum spacetime.  Of course, the most unorthodox part of this was the nonlocality of the  transfer of information from inside the horizon.
 -- I think many of my colleagues thought I was crazy to propose this, and I wondered if I indeed was.  But, I couldn't see an easier way. 
Since then,
it has been gratifying to see 
a number of other authors have ultimately come around to a similar viewpoint.  In particular, it is nice to see the amount of agreement with 
Carlo; as I show in 
Figure~\ref{SGfig1}, 
if one redraws his picture it clearly looks like such a massive remnant scenario.  Of course, in saying this, 
I am assuming 
that we should declare victory over the crisis once we see how the massive remnant forms; the puzzling attachment of the upper half of the picture, where one transitions to something more like a white hole, is less clearly necessary.  Likewise, proposals such as gravastars~\cite{Mazur:2004} and fuzzballs~\cite{fuzz} (at least in certain versions), and firewalls, appear to invoke the same crazy proposal -- information must transfer in a way that appears spacelike  when described with respect to the semiclassical geometry, in order to form a new quantum ``object" with the information at or outside the would-be horizon.

But, perhaps due to advancing age, I've become more conservative -- now I'm suggesting
that there could be a less-violent, but still ``nonlocal" (with respect to the semiclassical geometry) transfer of information, that can for example save the infalling observer from an untimely death.  Whatever the correct picture is, and given our uncertainty about the more fundamental dynamics, we should try to parameterize and constrain the relevant physics.  One relevant parameter is the time scale on which LQFT and semiclassical spacetime break down and  new dynamics manifests itself; we know that this must happen by a time $\sim R_S^3$, and should not happen before a time $\sim R_S\log R_S$. {The longer time scale is that on which the black hole appreciably shrinks, and by which information must start to emerge for a unitary description; the latter time scale is inferred from gedanken experiments where an outside observer collects information and then enters the black hole to try to compare with information inside the black hole.}  We can also parameterize the range over which the dynamics extends, and the ``hardness" ({\it e.g.} characteristic momentum scales) of the new dynamics.  All of these characteristics are important to describe what happens to an infalling observer, and also to describe possible observational signatures of such quantum modifications to black hole structure.

I'll close this {section}
with a final note about departures from the standard locality of LQFT.  As we know, in LQFT locality and causality are intimately linked; if  an observer can send a signal to a spacelike separated event, {we} can perform a Lorentz transformation and find that signal propagates back in time.  And, by combining two such signals, the observer can send a signal into their own backward lightcone, clearly violating causality -- and causing paradoxes.  Indeed,
in his comments at the meeting, 
Gerard (see Section~\ref{gerard})
indicated {\rm there is reluctance in the community} to give up locality and causality, possibly having this argument in mind.

The interesting thing is that if there are departures from standard locality that only really become relevant in the black hole context, these don't necessarily lead to anything that an observer would describe as acausality~\cite{NLvC}.  Specifically, information can fall into a black hole, and then be transmitted out at a later time, without any observer seeing a violation of causality.  The loophole in the preceding argument arises because the background black hole spacetime specifies a definite frame of reference, so the argument about constructing a signal propagating into your past light cone can't be made.  Moreover, with such a delay, with respect to the black hole's frame, one can satisfy an alternate macroscopic test of causality, which requires that in scattering outgoing signals should not precede the corresponding ingoing ones.  Thus, despite such nonlocality, it is still perfectly possible for the S-matrix to have the expected time delays~\cite{Erice,GiPo}.

\subsection{Violence is not the answer?}

Despite the fact that this 
{session} has the underlying theme of a ``duel'',
I'm going to
seek a nonviolent alternative -- for the physics!  Specifically, I propose 
to investigate interactions that represent a ``minimal" departure from the usual LQFT description of the immediate surroundings of a black hole.  This suggests that we work within an effective field theory framework, but consider adding new interaction terms that couple the black hole quantum state to the quantum fields in the immediate vicinity of the black hole~\cite{SGeft,GiShtwo}.  

In this spirit of minimality, I'll 
assume that these new interactions only extend over a range of size $\sim R_S$ outside the black hole horizon.  An important question is which LQFT modes they couple to.  For example, one could model a firewall by introducing interactions transferring information~\cite{Trieste} to very short wavelength (as seen by the infalling observer) modes that are very near the horizon.  A more benign scenario is, apparently, to transfer information to modes whose wavelength grows with the size $R_S$ of the black hole; a simplest example is to simply take this wavelength to be $\sim R_S$.

A priori, these interactions could transfer information (and correspondingly energy) to any of the quantum fields in nature~\cite{SGeft,GiShtwo}; for example just to fermions or just to gauge bosons. However, generic such interactions would spoil a particularly beautiful part of our current account of black holes:  black holes would no longer obey the laws of black hole thermodynamics.  It may be that there is no absolute necessity that they must respect these laws~\cite{SGstatph}, with the entropy formula given by Bekenstein and Hawking, but it seems desirable to preserve, at least approximately, such a nice story.

One reason generic interactions spoil black hole thermodynamics is because they would provide channels for information, and thus energy, to escape a black hole that are not universal; then, a black hole could not be brought into equilibrium with a thermal bath at the Hawking temperature -- detailed balance would fail.  An obvious way to achieve such universality is if the couplings to the black hole internal states are through the stress tensor~\cite{SGstress}.  Such couplings have another feature -- they make a story of information flow from a black hole more robust to ``mining" scenarios, where one introduces a cosmic string, or other mining apparatus, to increase the evaporation rate~\cite{Frolov:2002qd,Frolov:2000kx,Lawrence:1993sg,UnWa}. 
Such interactions through the stress tensor would also universally couple to the mining apparatus.  One might also be concerned that these stress-tensor  couplings would violate the Bekenstein-Hawking formula {because they increase} the energy flux from a black hole.  But, recall that emission from a black body is given by the Stefan-Boltzmann law, and in particular is proportional to the area of the emitting surface; in the black hole context, the new interactions may effectively be increasing that area~\cite{SGSB}.  

By following such a ``conservative" path, we have greatly limited our alternatives.  The new interactions universally couple to soft modes (wavelength $\sim R_S$)  in the immediate vicinity of the black hole, and the number of such modes that escape the black hole is limited.  And, these interactions have a job to do:  to unitarize black hole decay, they must transfer of order one qubit of information per time $R_S$, so that the entanglement entropy of the black hole with its surroundings decreases to zero by the time it has disappeared.  This constrains the strength of the new couplings; the simplest way to achieve such information transfer is if they provide an order one correction to the Hawking radiation. 

Thus, a rather simple and ``conservative" set of assumptions has led us to an interesting conclusion.  First, note that black hole state-dependent couplings to the stress tensor can be though of as a black hole state-dependent modification of the metric in the vicinity of the black hole.  Then, if such a coupling is of order unity, that corresponds to a metric deviation that is of order unity.  So, while the fluctuations in the ``effective metric" can be soft {(long wavelength)}, they are strong.  One might be initially concerned that such fluctuations would greatly alter the experience of the infalling observer.  However, with the softness scale set by $R_S$, typical curvatures measured by an infalling observer are only of size ${R}\sim 1/R_S^2$ -- the same size such an observer would see in Schwarzschild/Kerr.  So, and this addresses another concern expressed by {Gerard} in Section \ref{gerard}, the infalling observer doesn't need to see a drastic near-horizon departure from the experience expected for infall into a classical black hole.  

Before turning to the question of observation, it is worth reviewing the assumptions we have made, to emphasize their simplicity and paucity.  We need to reconcile black hole disintegration with quantum mechanics.  Assuming information can be localized to begin with, this implies that information must transfer from black hole states to outgoing radiation, and at a certain rate, of rough size one qubit per time $R_S$.  This can be accomplished with new quantum interactions.  To avoid a violent departure from the semiclassical picture, these interactions should couple to external modes with large wavelengths; the relevant scale could be $R_S^p$ for some $p>0$, but a simplest assumption is $p=1$.  We also assume that such new interactions don't extend beyond the immediate vicinity of the black hole, {that extends to a distance $\sim R_S$ from the horizon.}  Next, if we want to preserve black hole thermodynamics and in particular (at least approximately) the Bekenstein-Hawking formula for the entropy, {and address mining,} these interactions should couple universally to all modes, through the stress tensor.  Thus, they behave like black hole state-dependent metric fluctuations.  Finally, the size of these perturbations is determined by the statement that they need to provide an ${\cal O}(1)$ alteration to the Hawking radiation, imprinting information at the necessary rate.

\subsection{Observation via EHT}

While it is clearly important to further sharpen the preceding arguments, the prospect that order one departures from the Schwarzschild (or Kerr) metric in the immediate vicinity of the horizon are present raises the extremely interesting question of direct observation.  

In short, there has been a growing realization that quantum modifications to Schwarzschild/Kerr are needed on scales of order the horizon size (not just near the singularity); we have now entered the era where observations can be made of the near-horizon geometry, and so we should seek to observe or constrain such quantum structure.

In particular, as we heard at the meeting~\cite{Silke}, the Event Horizon Telescope (EHT) is now probing the structure of Sgr A* on the event horizon scale.  What might we look for, in a story where there is new quantum structure?  A preliminary discussion of this was given in \cite{SGobs}:  near-horizon perturbations in the effective metric seen by matter would lead to deviations from the geodesics predicted for the Schwarzschild/Kerr solutions.  This, in turn, would alter the  observed images, and specifically could alter the shape of the black hole shadow and photon ring that is seen just outside it.  A careful treatment of the possible effects on EHT images involves modeling the accretion flow -- which emits the light that EHT observes -- and numerical ray tracing to see the effects of the perturbations on the collection of light trajectories.  Such a treatment is in progress~\cite{GiPs}.  But, as I
pointed out in the discussion, generic expectations from such perturbations are a smaller, fuzzier shadow, and distortion of the photon ring.  {(Since this meeting and the original version of this writeup, this analysis has appeared \cite{GiPs2}.)}

\subsection{Post-LIGO update}\label{postLIGO}

Observation of gravitational radiation from inspiraling black holes also provides a possible probe of black hole structure, as was pointed out in \cite{SGobs} and preliminarily explored in \cite{SGGW}.  In short, if black holes have quantum structure not described by Schwarzschild/Kerr, then this is expected to alter their dynamics, particularly in the plunge/merger phase.  No large deviations were seen in the first LIGO detection~\cite{ligo}, which already indicates constraints on such scenarios.  Providing sharp constraints requires modeling the effects on the gravity wave signal due  to quantum black hole structure, which in particular probably requires treatment via numerical relativity, given the strong nonlinearities.  Improvement of LIGO sensitivity to possible deviations combined with further theoretical work thus offers the exciting possibility of learning more about such effects.  One might expect that ultimately the photon observations of EHT can more cleanly resolve deviations from general relativity, but LIGO may also ultimately offer the advantage of statistics of multiple events, so that remains to be seen.\footnote{For a different point of view on this subject, see the footnote at the end of section \ref{gerard}.}

\section{A Resolution to the Duel: Finding common ground in quantum gravity research\\ \small{\it Contribution by G.~'t Hooft}} \label{gerard}

The central question we face is how black holes can be properly incorporated in a grand scheme of quantum gravity. \\[5pt]
It is of tantamount importance to hold on as much as possible to the principles and symmetries that we are dealing with in today's theories. Starting from what we know, black holes will then display quantum  features that seem to be physically quite acceptable: they radiate particles of all sorts, with, on the average, thermal spectra. These properties appear to obey the usual conditions of causality and locality. An observer will not see any violations of these principles. Yet, when we try to find good descriptions of these phenomena, it seems some of us are inclined to throw locality and causality overboard, replace pure states by mixed states, and so on -- sometimes a bit too easily. A classical observer should still think all (s)he sees agrees with the classical theories.

On the other hand, could it be that a more `primitive' theory does not have general invariance -- or even quantum mechanics? In this respect it would be interesting to cite an observation by a computer scientist: as far as we know, all classical solutions of Einstein's equations have the property that signals are \emph{slowed down}, in comparison with any featureless vacuum solution. The computer scientist commented that this would be what one should expect if nature would be modelled as a computer: near a heavy body, this computer must process more information inside a smaller volume, so this goes slower.

In such a world view~\cite{thooft,'tHooft:1984aa,'tHooft:1984re,'tHooft:1986ub,'tHooft:1990fr,'tHooft:1992zk,'tHooft:1996tq}, 
what we call quantum mechanics today, may eventually become more like what thermodynamics is now: a description of some deterministic system in a statistical language. Note: thermodynamics can still be applied to single atoms, and similarly, quantum mechanics will still be true for small-mass black holes, but it could be that a deterministic theory will explain where these laws come from. That theory may turn out to be as deterministic as one's laptop.

Ideally, what should come out of an advanced theory is that, in the very end, black holes act just as heavy radioactive nuclei, or objects like a bucket of water: they absorb and emit particles, and, in practice, one can't distinguish whether they are in a pure quantum state or a mixed state. We should be able to describe them accurately. This means that, just as a bucket of water, one should be able to describe them as pure quantum states. Twenty years ago, this view, coming from particle physicists, was not very popular among the traditional practitioners of general relativity. 

There will be entangled components in these states. In particle physics, the lowest energy states are very well distinguishable. In the early days, the 1960s, complete lists of all particles and their properties were published yearly: the so-called Rosenfeld tables. Nowadays, such tables consist of thousands of pages. In principle, we should be able to repeat this for all black hole states, after which we should be able to decipher Nature's book keeping system.

Of particular interest would be black holes whose mass is 10 times to 1000 times the Planck mass, in the transition region between pure quantum objects and classical objects 
Note that, even if the size of such black holes is much smaller than the size scales in the Standard Model, they may still be in the classical regime. This is because, as soon as velocities approach that of light, the momentum $p$ of such a black hole can easily become much more than the Planck momentum. Therefore, the product of the uncertainties $\Delta p$ and $\Delta x$ will be much greater than $\hbar$. Therefore, black holes considerably heavier and bigger than the Planckian dimensions, may in all respects be handled as classical objects. For them, Einstein's equations make perfectly sense. The most challenging problem of black holes in physics is the regime where black holes, and the particles they interact with, all reside in, or near, the Planckian domain. 
Classical black holes obey a ``no-hair" theorem, but the quantum black hole has hair. One can see this if there is a scalar field present. At the horizon, such a field does not change under a Lorentz transformation, so its values on the horizon are conserved in time for the outside observer.

It is not true that a typical black hole horizon can only be defined if one waits for very long time periods.  As soon as a time of the order of \(m\log m\) has gone by,  in Planck units, one sees the typical horizon features such as Hawking radiation. That is a robust property of a horizon, and this amount of time is short. In all black hole theories I am considering, time scales are as short as \(m\log m\). We then have the internal degrees of freedom of the black hole, and those of particles surrounding it.

What one finds, in general, is that the gravitational back reaction of particles can in principle impart their information onto the out-going Hawking particles. The starting point is the Aichelburg-Sexl solution of the gravitational field of a fast moving particle~\cite{Aichelburg:1970dh,Dray:1984ha}. 
It is a common misconception that this effect will be small and unimportant. To the contrary, it diverges exponentially with time delay, such that, at time scales of the order of, and beyond, $m \log m$ this effect becomes dominant, and can completely explain how information re-emerges from a black hole. 
The price one pays is, that particles at the Planck scale cannot have the quantum numbers we are used to in the Standard Model: \emph{only} the geometric properties of particles should suffice to describe the information they carry.  This is because they return information by the way they interact with gravity. This is indeed as it is in string theories. Globally conserved quantum numbers such as baryon number will be untenable, since we know that a black hole can be formed from much more baryons than whatever can be returned when they decay.
  
As a note added in proof: in a recent investigation~\cite{Hooft:2016itl,Hooft:2016cpw},  a clearer picture is drawn of how all information can reappear, which is by a topological twist in space-time, a twist that cannot be observed directly by outside observers. It is a necessary result of the fact that the Penrose diagram of a stationary black hole (a hole that is much older than \(m\log m\) in Planck units) has \emph{two} asymptotic regions; one must have a good understanding of what they both mean. 
As soon as we accept the notion that particles in the Planck regime only have geometrical properties, the method I have been advocating for years, does away with the firewalls. This is because it gives a unitary evolution law just as the Schroedinger equation does, and the solutions of this law usually consist of entangled states. The gravitational drag effect, the central engine in this domain,  forces us to rephrase these states as soon as the time delays surpass $m \log m$, which is where the older, more primitive arguments would give us firewalls. The firewalls are transformed away now, as I've tried to explain in my most recent papers.

Yet other problems remain.
The total set of quantum states available for a black hole is discrete, all of space and time here seems to be discrete, and therefore we will have to give up some of our sacred notions that only make sense in a continuum, such as strict locality. Even the notion of probability will have to be reconsidered. The problem is not how to imagine crazy scenarios, the problem is how to arrive at the correct scenario by making only small steps, without having to make outrageous assumptions. There is no lack of information: we have special and general relativity, and the entire Standard Model of the elementary particles with some 25 parameters, coupling strengths that we do not know yet how to derive, but which can be measured accurately.

The Standard Model now, is the prototype of a successful development in science. It is based on Fock space approaches. In black hole physics, as in gravity in general, Fock space will not be good enough, so that I think gravity, and black holes, will put new constraints on the Standard Model. A theory better than Fock space may help us understand, and calculate, those 25+ parameters.

This information is notoriously difficult to implement, however. All today's theories are based on \emph{real numbers}: positions, momenta, energies, as well as constants of nature. But every single real number carries an infinity of information, and that's probably more than can be accommodated for inside volumes as small as the Planck volume. How do we make a theory without real numbers? Even the set of \emph{integers} might be too large. Should we exclusively employ bits and bytes?

Note, that eventually, when working out any theory, real numbers such as \(\pi\) and \(e\) \emph{will} show up soon enough, but only when you integrate things over larger volumes and distance scales.

How exactly to incorporate any most useful version of locality will be a difficult problem. In particular,  locality in the quantum mechanical sense is an issue to which special meetings are devoted.\footnote{
Regarding observational prospects and in direct response to Steve (see \ref{postLIGO}), the by far most likely scenario is that quantum effects will leave no trace in the behavior of kilometer-sized black holes, since we expect, like everywhere else in quantum mechanics, that all phenomena where the length scale, the time scale and the mass scale are way beyond the quantum regime, will be described by classical laws. In this case, these will be Einstein's equations, so that no deviations from the standard GR results will be seen to occur.
}

\section{Summary and Observational Prospects}

When brought to a final showdown, the Good, the Bad, and the Ugly of gravity and information do  leave room for non-violent resolution and redemption.
We witness a congregation of representatives from different prominent approaches, converging in similar ideas. Most notably, theoretical problems with local quantum field theory (LQFT), as well as those concerning current concepts of locality 
in general, are acknowledged across various communities. When venturing into the uncharted territory in the Wild West of our current knowledge, revisions of these concepts are considered foundational to the construction of quantum gravity.
Further consideration is required to demonstrate which technique will lead to a resolution of the current stand-off:
On one side, loop quantum gravity claims that the standard causal structure of spacetime fluctuates, leading to a necessary revision of how causality is defined. This is based on the idea that, according to general relativity (GR), the geometry of spacetime is determined by the gravitational field, which is a quantum field, undergoing quantum fluctuations and non-perturbative quantum effects.
On the other side, {a convincing argument has not been given that {\it short distance} fluctuations in the causal structure leads to the needed {\it long distance} modifications to a picture based on semiclassical spacetime.  So, the need for unitary evolution suggests that new non-local physical effects are needed.  This physics may be of a ``violent" nature, such as with massive remnants or firewalls, or of a ``non-violent" nature as with soft quantum structure of black holes.  }
Both sides seem to agree that significant modifications to mere perturbative quantization of matter and metric fluctuations on a background geometry are necessary. 
There is hope that these modifications are non-violent and could be added systematically as corrections to known physics. 
Effective field theory descriptions are proposed as a systematic and also the most conservative approach in order to capture the leading quantum gravity corrections to our classical gravity understanding.
One obvious example of such an effective description may be to take into account the backreaction of the Hawking radiation and its interaction with the infalling matter~\cite{Kiem:1995iy,thooft,Verlinde:2012cy}.
Potentially, we could also keep locality, causality, and the concept of everything being a pure state (even black holes from their formation until their evaporation) intact. This may require a re-interpretation of quantum mechanics (and LQFT) as an effective description, similar to statistical mechanics. In this analogy, black holes are like  buckets of water: we could try to follow all its constituents through the causal local evolution, but describing the system in the analog of a statistical ensemble would be much more appropriate. In such a description, globally conserved quantities (baryon number) seem untenable and unnecessary, because only the geometric properties (mass, energy-momentum tensor, topological information) of particles should matter to gravity, dictating the symmetries for a putative effective field theory description of quantum black holes and quantum gravity.
Taking an entirely different route, AdS/CFT may provide a fundamental formulation for quantum gravity, mapping a strongly curved regime of a gravitational theory to a weakly coupled field theory that can be described perturbatively.

And so, we reach the end of our story detailing the Good, the Bad, and the Ugly of gravitation and information with good news.  Currently, we are standing at the precipice of the strong gravity regime. Over the next decade we will open the window and gaze in.  As of February 2016, the LIGO Scientific Collaboration has confirmed the observation of gravitational waves due to merging black hole binaries, giving a first real test of general relativity in this limit \cite{ligo}.  In the not too distant future, the Event Horizon Telescope promises to provide the first image of a black hole's horizon at an incredible  resolution.  
On the theorists' wish list there are also 
observations of high curvature effects, long-time pile up of perturbative corrections, or obviously direct observation of deviations of the near-horizon geometry from Schwarzschild or Kerr geometry.
Also of interest is the possibility that the Event Horizon Telescope (EHT) may discover altered (non-GR) photon rings, and {distorted} shadows compared to GR.
LIGO may observe altered dynamics during the plunge or merger phase, or maybe a quasinormal mode ringdown that deviates from the one predicted by GR.

In parallel to this, there is still a great expectation for the results the LHC will provide in future runs. 
In the presence of extra dimensions \cite{Antoniadis:1998ig,ArkaniHamed:1998rs,ArkaniHamed:1998nn,Randall:1999ee,Randall:1999vf} 
gravitational collapse of colliding particles to form a microscopic (evaporating) black hole is possible at LHC working energies \cite{Argyres:1998qn,Banks:1999gd,Dimopoulos:2001hw,Giddings:2001bu}. 
Other particle physics testbeds include the formation of microscopic black holes in ultra high energy cosmic ray showers that might be observed at the Pierre Auger Observatory \cite{Feng:2001ib} or in AMANDA/IceCube, ANTARES neutrino telescopes \cite{AlvarezMuniz:2002ga,Kowalski:2002gb}. Interestingly non-classical effects might drastically change the production rates \cite{Mureika:2011hg} and the signature in detectors \cite{Gingrich:2010ed,Nicolini:2011nz,Rizzo:2006zb}. 
The above experiments aim to the observation of the evaporation end-point  of a microscopic black hole, a fact that {could} disclose crucial insights about how information is ultimately exchanged and preserved beyond what is known on the ground of standard semiclassical arguments. 

 The gravitational physics community looks forward to these, and future discoveries, with tempered anticipation, as LIGO, EHT and the LHC are staking new claims, eager to uncover observational gold mines. 
   






\end{document}